\documentclass[9pt,conference]{IEEEtran}

\usepackage{waspaa25} 

\usepackage{cite}

\usepackage{bm} 
\usepackage{adjustbox} 
\usepackage{caption}

\title{Context-Aware Query Refinement for Target Sound Extraction: \\  Handling Partially Matched Queries}

\name{Ryo Sato$^{1}$,
      Chiho Haruta$^{1}$,
      Nobuhiko Hiruma$^{1}$,
      Keisuke Imoto$^{2}$
      }
\address{$^{1}$RION Co., Ltd., Tokyo, Japan\;
$^{2}$Kyoto University, Kyoto, Japan\\
}

\begin{document}

\maketitle

\begin{abstract}
Target sound extraction (TSE) is the task of extracting a target sound specified by a query from an audio mixture. Much prior research has focused on the problem setting under the Fully Matched Query (FMQ) condition, where the query specifies only active sounds present in the mixture. However, in real-world scenarios, queries may include inactive sounds that are not present in the mixture. This leads to scenarios such as the Fully Unmatched Query (FUQ) condition, where only inactive sounds are specified in the query, and the Partially Matched Query (PMQ) condition, where both active and inactive sounds are specified. Among these conditions, the performance degradation under the PMQ condition has been largely overlooked. To achieve robust TSE under the PMQ condition, we propose context-aware query refinement. This method eliminates inactive classes from the query during inference based on the estimated sound class activity. Experimental results demonstrate that while conventional methods suffer from performance degradation under the PMQ condition, the proposed method effectively mitigates this degradation and achieves high robustness under diverse query conditions.
\end{abstract}

\section{Introduction}
\label{sec:intro}

Target sound extraction (TSE) is the task of extracting one or more target sources from a mixture, specified by auxiliary information known as a query (or clue/hint)~\cite{ochiai2020listen,delcroix2022soundbeam}. TSE has potential applications in hearing aids, telephony, and environmental sound monitoring. Various formats of queries have been explored, including predefined class labels~\cite{ochiai2020listen}, audio samples~\cite{delcroix2022soundbeam}, and text descriptions~\cite{kilgour2022text,liu2022separate}.

Much of the prior research on TSE implicitly assumes that all sound sources in the mixture are known~\cite{ochiai2020listen,liu2022separate,kilgour2022text,veluri2023real,baligar2024catse,wang2022improving,kamo2023target,hai2024dpm,liu2024separate,wakayama2024online,hernandez2025soundbeam,kim2024acoustic,wu2025cross,hernandez2024interaural,veluri2023semantic,ma2024clapsep,kim2024improving,wang2025soloaudio,wang2024tse,choi2025multichannel}. Under this assumption, the query specifies only the active sources present in the mixture as target sources. In this study, we refer to this as the Fully Matched Query (FMQ) condition. However, in realistic usage such as setting a query while listening through a hearable device, it is difficult for users to perfectly identify all sources present. Users must often guess the active sources, and mistakes may lead to inactive sources being included in the query. 

Some other studies address the Fully Unmatched Query (FUQ) condition where all target sound sources specified in the query are inactive in the mixture. In this case, an ideal TSE system should output silence. A training method using inactive samples (IS) has been proposed to address this condition, where IS represents samples in which the specified target sound source is absent from the mixture~\cite{delcroix2022soundbeam}. While this approach can bring the output closer to a zero signal under FUQ conditions, it has been reported to involve a trade-off and degrade performance under FMQ conditions. Researchers have also considered another approach involving methods that perform target sound detection separately from TSE to replace the output signal with a zero signal. These do not degrade the performance under FMQ condition but are limited to single-class extraction~\cite{baligar2022cossd,baligar2024mcrtse}.

To make the problem setting more realistic, it is necessary to consider scenarios where multiple target sounds are specified in the query, but only some of them are active in the mixture, while the rest are inactive. In this study, we refer to this as the Partially Matched Query (PMQ) condition. In this case, an ideal TSE system is required to ignore the inactive classes specified in the query and extract only the active sources. However, specifying inactive target sounds in the query increases the risk of performance degradation due to the erroneous extraction of non-target sounds. The training method with IS is unlikely to be effective in preventing such performance degradation. Furthermore, methods based on target sound detection can only replace the output signal with a zero signal and are fundamentally unable to handle the PMQ condition. To the best of our knowledge, the adverse effects of the PMQ condition on performance have been overlooked till now. This study is the first to focus on this problem, clarifying its severity and proposing a solution.

This research aims to develop a novel method that operates robustly under the PMQ conditions. To achieve this goal, we propose context-aware query refinement that estimates the sound class activity in the mixture during inference and refines the original query by removing inactive classes. This aims to extract only the active target sounds and prevent performance degradation caused by inactive classes included in the query. As sound class estimation and TSE are closely related tasks, we efficiently implement the proposed method by training a shared feature extractor through multi-task learning.

Our experimental results demonstrate that conventional TSE methods suffer significant performance degradation under PMQ conditions, whereas the proposed method effectively mitigates such degradation, achieving high robustness under diverse query conditions. This approach not only improves the robustness under PMQ conditions, but also handles FUQ conditions, thereby enhancing the robustness under more realistic and varied query scenarios.

\begin{figure*}
  \centering
  \vspace{-4pt}
  \includegraphics[width=0.99\linewidth]{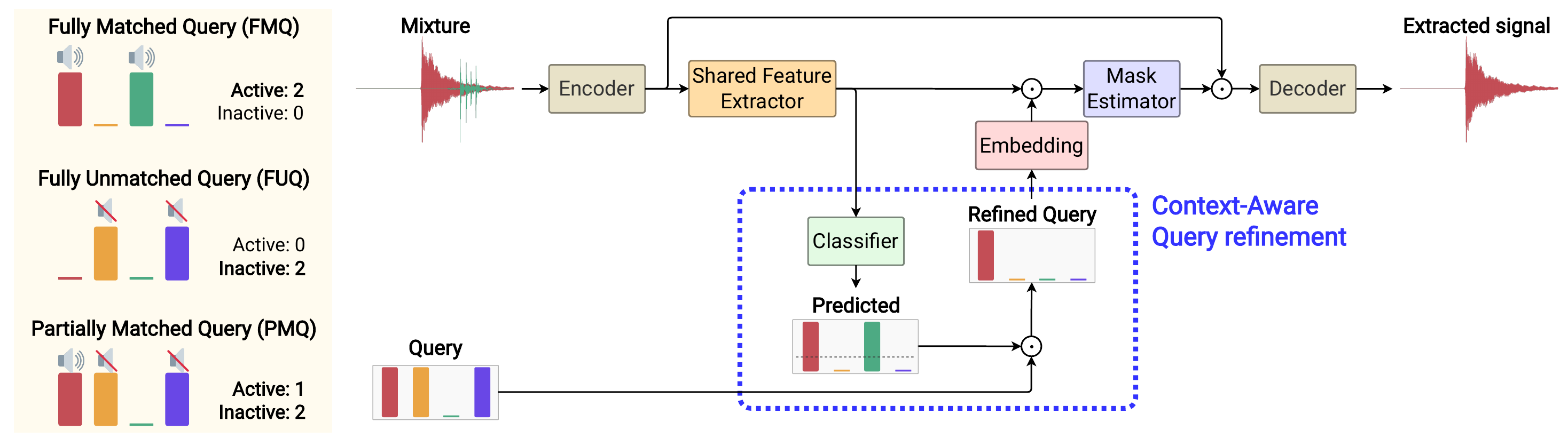}
  \caption{Overall architecture of the proposed query refinement method. Example queries for each condition are shown on the left.}
  \vspace{-4pt}
  \label{fig:proposed_method}
\end{figure*}

The main contributions of this study are two-fold:
\begin{itemize}
    \item We identify and highlight the performance degradation problem under the Partially Matched Query (PMQ) condition in TSE, an issue that is practically important, but has been largely overlooked.
    \item We propose context-aware query refinement, which modifies the query based on the estimated sound class presence in the mixture, and show that it effectively suppresses performance degradation under PMQ conditions.
\end{itemize}

\section{Related Work}
\label{sec:related_work}

\subsection{Target speaker extraction}
Target speaker extraction is closely related to our work, as it also addresses scenarios where the target speaker may or may not be present in the audio mixture~\cite{zhang2020x,borsdorf2021universal,zhang2021towards,delcroix2022listen,zhang2023speaker}.

Methods have been proposed to suppress the extraction of non-target speakers by training models with samples where the target speaker is absent~\cite{zhang2020x,borsdorf2021universal}. However, similar to the training method with IS~\cite{delcroix2022soundbeam}, such methods involve a trade-off between suppressing incorrect extractions and maintaining the extraction performance for the actual target speaker.

Other studies involve introducing an additional speaker verification module to replace the output signal with silence when the target speaker is absent~\cite{zhang2021towards,delcroix2022listen,zhang2023speaker}. These approaches allow handling the absent target condition without degrading the extraction performance, as the model is trained only on present target conditions. Typically, these approaches generally assume single-speaker extraction, where it is sufficient to simply replace the output with silence. However, the PMQ conditions considered in our study require extracting only active classes while avoiding non-target sounds, thereby making these approaches not directly applicable.

\subsection{Target sound extraction}
To address the FUQ condition in TSE, a training method with IS was proposed~\cite{delcroix2022soundbeam}. This method yielded outputs closer to a zero signal under FUQ conditions, but as with target speaker extraction, performance degradation under FMQ conditions remains a challenge. Furthermore, the PMQ condition, which is the focus of our study, was not considered, indicating that realistic scenario settings were not sufficiently explored.

\section{Proposed Method}
\label{sec:method}

To realize a robust TSE system under PMQ conditions, we propose context-aware query refinement. The overall architecture is shown in ~\cref{fig:proposed_method}. The core idea is to utilize the estimated results of sound class activity during inference to eliminate inactive classes in the query. By using only FMQ conditions during training, we aim to maximize the model's extraction performance, while mitigating performance degradation under PMQ conditions during inference. This section details the architecture, query refinement process, and training method.

\subsection{Architecture}
\label{ssec:arc}

To efficiently implement the proposed query refinement, we adopt an architecture that jointly performs TSE and sound class estimation. This architecture consists of learnable encoder and decoder modules, a shared feature extractor for both tasks, and task-specific modules (a mask estimator and a classifier). The basic design of the TSE architecture is based on 
~\cite{ochiai2020listen}.
The details are described below.

\noindent\textbf{Encoder}:
The input mixture $\boldsymbol{x} \in \mathbb{R}^T$ is transformed into a feature representation $\boldsymbol{X} \in \mathbb{R}^{D \times L}$ by a learnable encoder:
\begin{equation}
    \label{eq:encoder}
    \boldsymbol{X}=\mathrm{Encoder}(\boldsymbol{x}),
\end{equation}
where $T$, $D$, and $L$ represent the number of samples in the input signal, the number of filters in the encoder, and the number of frames in the feature representation, respectively. The encoder consists of 1-D convolutional layers.

\noindent\textbf{Shared feature extractor}:
The shared feature extractor extracts shared features $\boldsymbol{Z} \in \mathbb{R}^{N \times L}$ used commonly by both TSE and sound class classification tasks from the mixture features $\boldsymbol{X}$:
\begin{equation}
    \label{eq:shared}
    \boldsymbol{Z}=\mathrm{f_{shared}}(\boldsymbol{X}),
\end{equation}
where $N$ and $\mathrm{f_{shared}}$ represent the feature dimension and the shared feature extractor, respectively. The shared feature extractor is composed of stacks of 1-D convolutional blocks, based on the architecture of the mask estimator in Conv-TasNet~\cite{luo2019conv}.

\noindent\textbf{Mask estimator}:
The mask estimator estimates a mask $\boldsymbol{M}\in\mathbb{R}^{D \times L}$ for extracting the target sound source based on the shared features $\boldsymbol{Z}$ and the query $\boldsymbol{q} \in \{0,1\}^{C}$ as follows,
\begin{equation}
    \label{eq:emb}
    \boldsymbol{e}=\mathrm{Embedding}(\boldsymbol{q}),
\end{equation}
\begin{equation}
    \label{eq:mask}
    \boldsymbol{M}=\mathrm{f_{mask}}(\boldsymbol{Z, e}),
\end{equation}
where $C$ and $\mathrm{f_{mask}}$ represent the total number of classes and the mask estimator, respectively. The query $\boldsymbol{q}$ is represented as a multi-hot vector indicating the target classes for extraction. It is transformed into an embedding vector $\boldsymbol{e}\in\mathbb{R}^N$ by the embedding layer and then conditioned on the shared features $\boldsymbol{Z}$ by element-wise multiplication. Similar to the shared feature extractor, the mask estimator is composed of stacks of 1-D convolutional blocks.

\noindent\textbf{Decoder}:
To extract the target sound from the mixture, the estimated mask $\boldsymbol{M}$ is applied to the mixture features $\boldsymbol{X}$, and the result is reconstructed into a time-domain signal by the decoder:
\begin{equation}
    \label{eq:decoder}
    \hat{\boldsymbol{s}} = \mathrm{Decoder}(\boldsymbol{X}\odot\boldsymbol{M}),
\end{equation}
where $\hat{\boldsymbol{s}}\in\mathbb{R}^T$ is the estimated target sound. The decoder consists of 1-D transposed convolutional layers.

\noindent\textbf{Classifier}:
The classifier estimates the existence probability $\hat{\boldsymbol{p}}\in[0,1]^C$ for each sound class in the mixture, based on the shared features $\boldsymbol{Z}$:
\begin{equation}
    \label{eq:cls}
    \hat{\boldsymbol{p}}=\mathrm{f_{cls}}(\boldsymbol{Z}),
\end{equation}
where $\mathrm{f_{cls}}$ represents the classifier. As this classifier uses the shared features $\boldsymbol{Z}$ as the input before conditioning, the estimated probability $\hat{\boldsymbol{p}}$ is determined only by the mixture and does not vary with the query.

The classification task performed by the classifier can be determined based on the dataset used and application constraints, with possibilities including audio tagging or sound event detection. In this work, as we use a dataset without frame-level event labels, we employ weakly-supervised sound event detection. The classifier consists of two BiGRU layers followed by a linear layer. Frame-level predictions are aggregated into clip-level predictions using a pooling function.

\subsection{Context-aware query refinement}
\label{ssec:cqr}
During inference, the proposed context-aware query refinement is performed using the query $\boldsymbol{q}$ and the estimated probability $\hat{\boldsymbol{p}}$ from the classifier:
\begin{equation}
    \label{eq:cqr}
    q_i^{\rm{refined}} =
        \begin{cases} 
            q_i & \text{if } \hat{p}_i \geq \theta \\
            0 & \text{otherwise},
        \end{cases}
\end{equation}
where $i$ is the index for the $i$-th class, and $\theta$ is the threshold for binarizing $\hat{p}_i$. In practice, the query refinement can be implemented by calculating the element-wise product between the binarized predictions and the original query.

If oracle classification results are used for query refinement, the PMQ conditions can be perfectly converted to the FMQ conditions, enabling the extraction of only the active target sounds within the mixture. This represents the upper bound performance of the proposed method under PMQ conditions. Furthermore, under the FUQ condition, the query is replaced by a zero vector, which is expected to bring the output closer to silence.

It should be noted that the performance of query refinement depends on the prediction accuracy of the classifier. The impacts of false positives and false negatives on query refinement performance are as follows:
\begin{itemize}
    \item \textbf{False positive:} A false positive occurs when an inactive class in the mixture is incorrectly predicted as active. If the original query does not include this class, the query refinement has no adverse effect. Conversely, if the query does include this class, the inactive class is retained in the query, and the refinement fails to provide improvement.
    \item \textbf{False negative:} A false negative occurs when an active class in the mixture is incorrectly predicted as inactive. If this occurs for a target class included in the original query, this target class is erroneously removed from the query by the refinement process, making extraction difficult. This is a potential risk of the proposed method.
\end{itemize}
Considering the above, it is important for the proposed method to minimize the occurrence of false negatives. Therefore, it is desirable to set the threshold $\theta$ to a relatively small value.

\subsection{Model training}
Assuming that context-aware query refinement functions ideally, the TSE model only needs to consider the FMQ condition. Therefore, we use only FMQ conditions during training.

The training loss function $\mathcal{L}$ is a weighted sum of the loss term for TSE, $\mathcal{L}_{\rm{tse}}$, and the loss term for sound class estimation, $\mathcal{L}_{\rm{cls}}$:
\begin{equation}
    \label{eq:loss}
    \mathcal{L}=\mathcal{L}_{\rm{tse}}+\lambda \mathcal{L}_{\rm{cls}},
\end{equation}
where $\lambda$ is a hyperparameter to balance the two tasks.

\begin{table*}[ht]
    \centering
    \vspace{-4pt}
    \caption{Experimental results under each condition. $\theta$ for the proposed method represents the query refinement threshold.}
    \sisetup{
    reset-text-series = false,
    text-series-to-math = true,
    mode=text,
    tight-spacing=true,
    round-mode=places,
    round-precision=2,
    table-format=2.2,
    table-number-alignment=center
    }
    \vspace{-4pt}
    \label{tab:exp}
    \captionsetup{justification=centering}
    \setlength{\tabcolsep}{4pt} 
    \resizebox{0.885\linewidth }{!}{
    \begin{tabular}{@{}l c S[table-format=2.2] 
                     S[table-format=2.2] S[table-format=2.2] 
                     S[table-format=-2.2] 
                     S[table-format=1.3] 
                     S[table-format=1.2] S[table-format=2.2]@{}} 
        \toprule
         &  & \multicolumn{1}{c}{\textbf{FMQ}} & \multicolumn{2}{c}{\textbf{PMQ}} & \multicolumn{1}{c}{\textbf{FUQ}} & \multicolumn{1}{c}{\textbf{Classification}} & & \\
        \cmidrule(lr){3-3} \cmidrule(lr){4-5} \cmidrule(lr){6-6} \cmidrule(lr){7-7}
         Method & IS & {SNRi (1:0) $\uparrow$} & {SNRi (1:1) $\uparrow$} & {SNRi (1:3) $\uparrow$} & {$\mathcal{A}^{\text{mix}}$ (0:1) $\downarrow$} & {Macro F1 $\uparrow$} & {MACs (G/s)} & {\# Params (M)} \\
         \midrule
         Baseline 1 & {-} & \bfseries 15.65 & 14.29 & 10.96 & -32.41 & {-} & 5.19 & 13.06 \\
         Baseline 2 & \checkmark & 14.71 & 14.56 & 12.66 & \bfseries -78.13 & {-} & 5.19 & 13.06 \\
         \midrule
         Proposed ($\theta=0.00$) & - & \bfseries 15.65 & 14.44 & 11.26 & -34.25 & {-} &  & \\
         Proposed ($\theta=0.05$) & - & 14.94 & \bfseries 14.65 & 14.21 & -48.54 & 0.635 &  &  \\
         Proposed ($\theta=0.10$) & - & 14.87 & 14.63 & 14.24 & -48.82 & 0.647 & 5.43 & 13.67 \\
         Proposed ($\theta=0.15$) & - & 14.84 & 14.60 & \bfseries 14.26 & -49.01 & 0.652 &  &  \\
         Proposed ($\theta=0.20$) & - & 14.80 & 14.58 & \bfseries 14.26 & -49.12 & 0.656 &  &  \\
         \bottomrule
    \end{tabular}
    }
\end{table*}

\section{Experiments}
\label{sec:experiments}

\subsection{Experimental setup}

\noindent\textbf{Dataset preparation}:
We used the FSDKaggle2018 dataset~\cite{eduardo2018} for foreground sounds and the TAU Urban Acoustic Scenes 2019 dataset~\cite{mesaros2018tau} for background sounds. The foreground sounds comprised 41 classes, a subset of the AudioSet ontology~\cite{audioset}. Mixtures were synthesized by combining 3-5 foreground sound classes with one background noise. The Signal-to-Noise Ratio (SNR) was randomly set between 15 and 25 dB. The duration of the synthesized mixtures was 6 seconds. The dataset was split into training (50k), validation (5k), and test (10k) sets. For computational efficiency, the sampling frequency was set to 16 kHz.

\noindent\textbf{Model architecture}:
The architecture of the proposed method is as described in  ~\cref{ssec:arc}. For the baseline method, we used a model derived from our proposed architecture by excluding the classifier module. The specific parameter settings for each component were as follows: For the encoder and decoder, the window length was 5 ms, the overlap was 50\%, and $D=256$. For the shared feature extractor and mask estimator, following the Conv-TasNet~\cite{luo2019conv} notation: $P=3$, $H=512$, $B=256$, $Sc=256$. The number of convolutional blocks per stack $X$ and the number of stacks $R$ were $X=8, R=1$ for the shared feature extractor, and $X=8, R=3$ for the mask estimator. The hidden dimension of the BiGRU layer in the classifier was 256. Frame-level predictions were aggregated into clip-level predictions using linear softmax pooling~\cite{wang2019comparison}.

\noindent\textbf{Training details}:
We used the Adam optimizer~\cite{kingma2015adam} with a batch size of 8 and trained for 100 epochs. The learning rate was warmed up linearly to 5e-4 over the first 10 epochs, followed by cosine annealing decay~\cite{loshchilov2017sgdr} to 0.

\noindent\textbf{Compared systems}:
To validate the effectiveness of the proposed method, we trained and evaluated the following three systems:
\begin{itemize}
    \item \textbf{Baseline 1}: The baseline architecture trained only under FMQ conditions.
    \item \textbf{Baseline 2}: The baseline architecture trained under both FMQ and FUQ conditions (using IS on 10\% of the training data).
    \item \textbf{Proposed}: The proposed model performing TSE and weakly-supervised sound event detection, trained only under FMQ conditions.
\end{itemize}

\noindent\textbf{Loss function}:
For the proposed method, we used the negative thresholded SNR ~\cite{MixIT} for $\mathcal{L}_{\rm{tse}}$ and binary cross-entropy for $\mathcal{L}_{\rm{cls}}$, performing multi-task learning with $\lambda=1$. Baseline 1 was trained using only $\mathcal{L}_{\rm{tse}}$. Baseline 2 was trained using the same loss function as \cite{delcroix2022soundbeam}, setting the target signal to zero signal for IS.

\noindent\textbf{Evaluation}:
To confirm the performance under various query conditions, we evaluated by varying the number of active target classes $n_{\text{active}}$ and inactive target classes $n_{\text{inactive}}$ in the query. In the following sections, the query setting for each condition is denoted as $(n_{\text{active}}:n_{\text{inactive}})$. For the PMQ and FUQ conditions, inactive classes were randomly added to the query for each sample.

For performance evaluation, we used the SNR improvement (SNRi) [dB] relative to the input mixture to evaluate the extraction performance for active classes under the FMQ and PMQ conditions. For the FUQ condition, following \cite{delcroix2022soundbeam}, we used the attenuation ratio between the mixture and the extracted signal $\mathcal{A}^\text{mix}$ [dB] defined as,
\begin{equation}
    \mathcal{A}^{\text{mix}} = - 10 \log_{10} \left( \frac{\| \boldsymbol{x} \|^{2}}{\| \hat{\boldsymbol{s}} \|^{2}} \right),
\end{equation}
to evaluate the closeness of the extracted signal to silence.

\subsection{Performance on FMQ condition}
We evaluated the performance of the proposed method under the FMQ condition without query refinement ($\theta=0.00$). As shown in \Cref{tab:exp}, the SNRi achieved by the proposed method was comparable to that of baseline 1 under the FMQ (1:0) condition. This suggests that the multi-task learning in our approach does not impair the performance on the primary TSE task. On the other hand, baseline 2 trained with IS exhibited a performance degradation, which is consistent with the results reported in \cite{delcroix2022soundbeam}.

\begin{figure}
  \centering
  \vspace{-4pt}
  \includegraphics[width=0.935\linewidth]{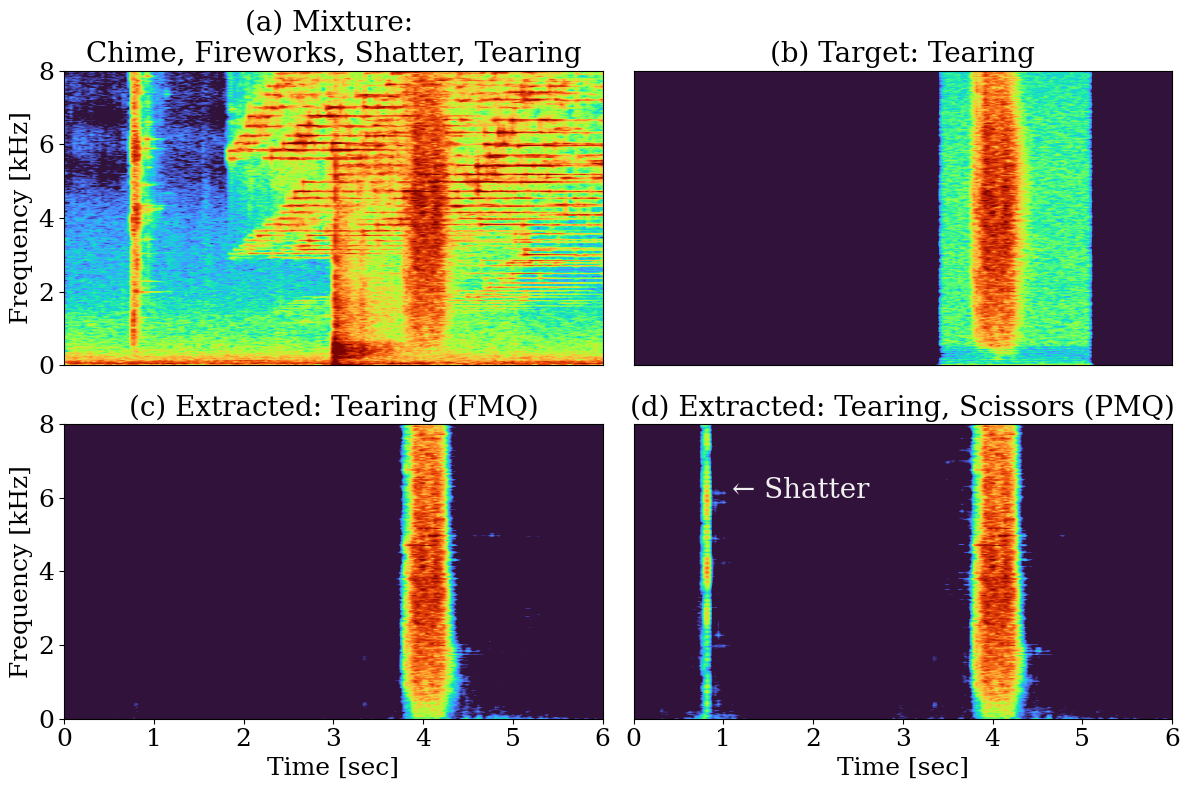}
  \vspace{-6pt}
  \caption{Example of performance degradation under the PMQ condition using baseline 1. (a) Mixture. (b) Target signal. (c) Ideal extraction result under FMQ condition. (d) Erroneous extraction under PMQ condition.}
  \label{fig:pmq_spec}
\end{figure}

\subsection{Performance on PMQ condition}
We demonstrate the performance degradation under the PMQ condition, which is the focus of this study. \Cref{fig:pmq_spec} shows an example of performance degradation with baseline 1 when an inactive class was included in the query under the PMQ condition. In this example, while the target class was ``Tearing'', the query mistakenly included ``Scissors'', an inactive class in the mixture. This resulted in the erroneous extraction of ``Shatter'', which was a non-target sound. Under PMQ conditions, severe performance degradation can occur in conventional TSE systems due to such erroneous extraction of non-target sounds.

\begin{figure}
    \centering
    \vspace{-4pt}
    \includegraphics[width=.975\linewidth]{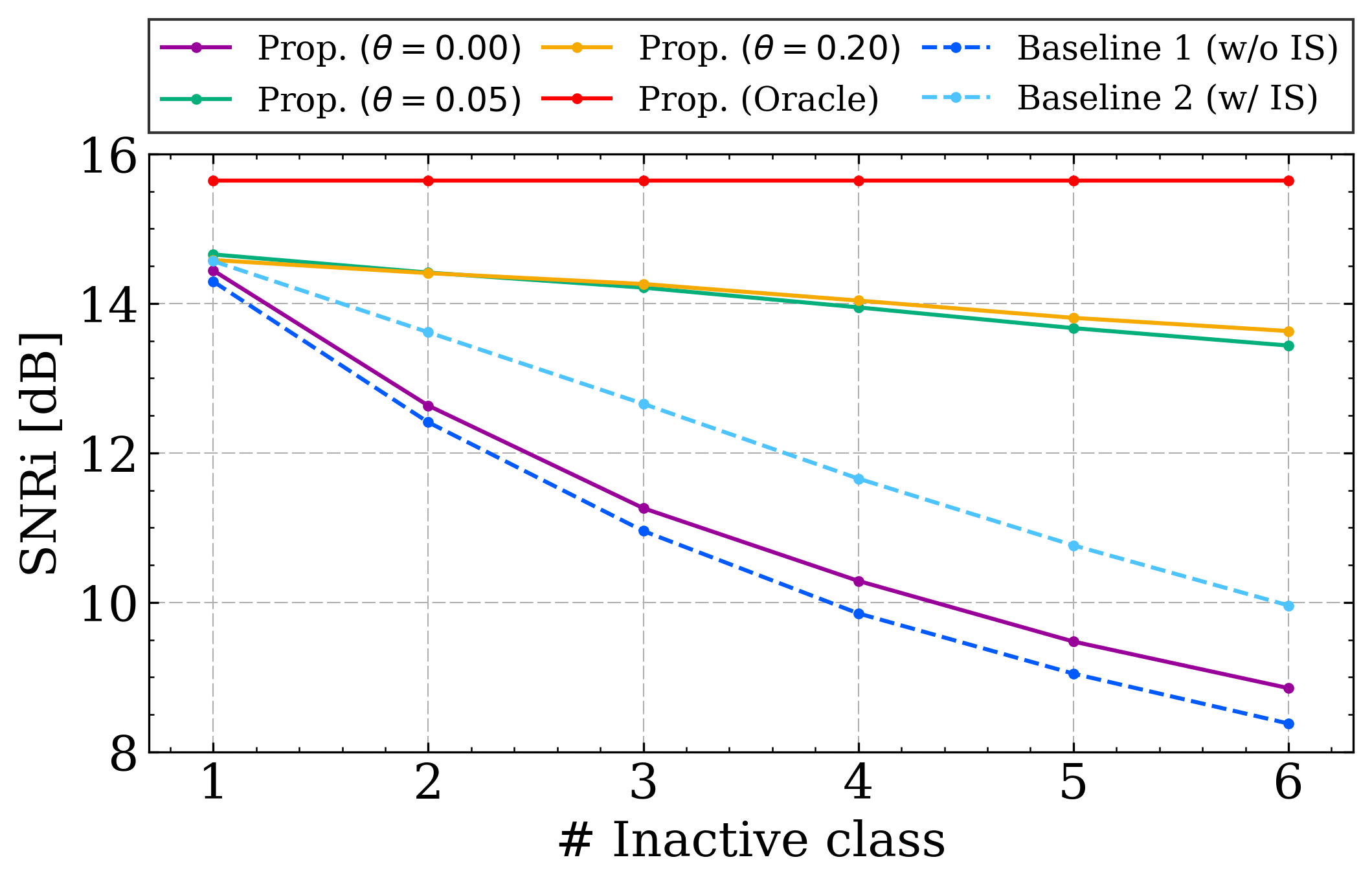}
    \vspace{-6pt}
    \caption{Relationship between $n_{\text{inactive}}$ and SNRi under PMQ conditions.}
    \label{fig:n_inactive}
\end{figure}

As shown in \cref{tab:exp}, the SNRi decreased for all baselines and the proposed method without query refinement ($\theta=0.00$) under the PMQ (1:1) and PMQ (1:3) conditions. Furthermore, \cref{fig:n_inactive} clearly illustrates that the performance degradation becomes sharply more severe for all methods with an increase in the number of inactive classes in the query. 

For the proposed query refinement method, we evaluated the effectiveness of mitigation of performance degradation under PMQ conditions. As shown in \cref{tab:exp}, the proposed method with query refinement reduced the performance degradation under the PMQ (1:1) and PMQ (1:3) conditions. Moreover, \cref{fig:n_inactive} shows that the proposed method with query refinement (green and orange lines) maintained high performance even with an increasing number of inactive classes, demonstrating a significant improvement in robustness under PMQ conditions. The performance using oracle classification results for query refinement (red line) indicates the potential for further performance gains by improving the classification accuracy.

\subsection{Performance on FUQ condition}

Under the FUQ condition, baseline 2 exhibited the best performance, demonstrating the effectiveness of training with IS for the FUQ condition, as shown in \cref{tab:exp}. As the proposed method did not consider the FUQ condition during training, its performance was inferior compared to baseline 2. However, applying query refinement yielded better performance compared to not applying it.
This improvement suggests that replacing the query with a zero vector brings an output closer to silence, as expected in \cref{ssec:cqr}.
The result indicates that the proposed method is useful not only for improving the performance under PMQ conditions, but also under FUQ conditions.

\subsection{Discussion of trade-offs and limitations}

While the proposed method offers the advantage of improving the performance under the PMQ and FUQ conditions, a trade-off exists with performance under the FMQ condition. As observed in \cref{tab:exp}, applying query refinement under the FMQ (1:0) condition leads to a lower performance compared to not applying it ($\theta=0.00$). As discussed in \cref{ssec:cqr}, this performance degradation is due to false negatives in the classification results. Such errors cause target classes that should be extracted to be excluded from the query, leading to extraction failure.

The proposed method emphasizes efficiency, limiting the increase in computational cost (MACs) and parameter size to only 4.7\% and 4.6\%, respectively, compared to the baseline. Consequently, the classifier does not achieve very high performance, with a Macro F1 score of approximately 0.65. Nevertheless, the experimental results demonstrate the effectiveness of query refinement under the PMQ and FUQ conditions. If the accuracy of the classifier can be improved, specifically reducing the false negative rate, it can potentially mitigate the performance degradation in the FMQ condition,  while further enhancing the robustness under the PMQ and FUQ conditions.





\section{Conclusion}
\label{sec:conclusion}

In this study, we addressed the performance degradation problem under the practically important PMQ conditions in TSE. We proposed context-aware query refinement using the estimated sound class activity to refine class label-based queries during inference through an efficient multi-task architecture. Experiments showed that our method effectively mitigates performance degradation under PMQ conditions. It also improves the performance under the FUQ condition. However, a trade-off exists between maintaining the FMQ performance and achieving robustness under PMQ conditions. Future work could involve leveraging temporal information from sound event detection for query refinement with higher temporal resolution, considering intra-clip PMQ conditions.

\clearpage
\bibliographystyle{IEEEtran}
\bibliography{refs25}

\end{document}